\documentclass[]{spie}  %>>> use for US letter paper
\usepackage{amsmath,amsfonts,amssymb}
\usepackage{aas_macros}
\usepackage{graphicx}
\usepackage{xcolor}
\usepackage[colorlinks=true, allcolors=blue]{hyperref}

\title{Towards quantum-enhanced long-baseline optical/near-IR interferometry}

\author[a]{Jayadev K. Rajagopal}
\author[a]{Ryan M. Lau}
\author[b]{Isack Padilla}
\author[a]{Stephen T. Ridgway}
\author[b]{Chaohan Cui}
\author[a]{Brittany McClinton}
\author[b]{Aqil Sajjad}
\author[a]{Stuartt Corder}
\author[a]{Mark Rawlings}
\author[a]{Fredrik Rantakyro}
\author[b]{J. Gabriel Richardson}
\author[b]{Amit Ashok}
\author[b]{Saikat Guha}

\affil[a]{NSF National Optical-Infrared Astronomy Research Lab., United States}
\affil[b]{The Univ. of Arizona, United States}

\authorinfo{Further author information: (Send correspondence to J.K.R.~or R.M.L.)\\J.K.R.: E-mail: jayadev.rajagopal@noirlab.edu, \\  R.M.L.: E-mail: ryan.lau@noirlab.edu}

\begin{document} 
\maketitle

\begin{abstract}
Microarcsecond resolutions afforded by an optical-NIR array with kilometer-baselines would enable breakthrough science. However significant technology barriers exist in transporting weakly coherent photon states over these distances: primarily photon loss and phase errors. Quantum telescopy, using entangled states to link spatially separated apertures, offers a possible solution to the loss of photons. We report on an initiative launched by NSF NOIRLab in collaboration with the Center for Quantum Networks and Arizona Quantum Initiative at the University of Arizona, Tucson, to explore these concepts further. A brief description of the quantum concepts and a possible technology roadmap towards a quantum-enhanced very long baseline optical-NIR interferometric array is presented. An on-sky demonstration of measuring spatial coherence of photons with apertures linked through the simplest Gottesman protocol over short baselines and with limited phase fluctuations is envisaged as the first step.
\end{abstract}


\section{INTRODUCTION}
\label{sec:intro}  

Optical and infrared long-baseline interferometry has emerged as a powerful technique allowing us to achieve angular resolutions down to milliarcseconds scales. This is notably highlighted by the scientific and technical progress in the past few years from interferometers with maximum baselines exceeding 200 meters such as the Georgia State University Center for High Angular Resolution Astronomy (CHARA) Array and the European Southern Observatory (ESO) Very Large Telescope Interferometer (VLTI). Optical and IR observations with milliarcsecond angular resolution are important for advancing a broad range of research areas in astronomy including stellar astrophysics, star and planet formation, and active galactic nuclei (See \citenum{Eisenhauer2023} for a review). 

The scientific promise of long-baseline optical and IR interferometry motivates an exploration of future capabilities beyond the current "classical" methods. Innovative applications of quantum-enhanced techniques\cite{Gottesman2012,Khab2019prl, Khab2019pra,BH2021} hold the potential to achieve very long ($\gtrsim1000$ m) baselines that can deliver \textit{micro}arcsecond-scale angular resolution at optical/IR wavelengths. Such high resolution could enable breakthrough science in a broad range astronomical research, especially exoplanets, stellar physics, and supermassive black holes.

\section{Performance Limits of ``Classical" Interferometry and the need for New Technology}
Among the challenges of optical interferometry, some may be addressed with the extension of current optical technologies. Sensitivity is perhaps most “easily” addressed by utilizing larger aperture telescopes. Dynamic range may be improved by teaming adaptive optics and coronagraphic technologies. Extension to much shorter wavelengths can be enabled by working in a space environment.

On the other hand, achieving greatly increased angular resolution with optical interferometry appears more problematic. Angular resolution is limited primarily by the available telescope separations (i.e.~the optical baselines).  These in turn are currently limited, first,  by the need to directly combine the collected wavefronts. To support this, it is necessary to transmit optical beams efficiently and without distortion over distances comparable to the telescope separations. Interferometer baselines can be increased significantly from the current values $<$1 km, to possibly several tens of km. Extension to 100’s of km, which is necessary for microarcsecond angular resolution at typical visible-near-infrared wavelengths, would be difficult and maybe impossible with conventional optical interferometry. Transmission losses, especially in the shorter wavelengths will overwhelm the signal.  Furthermore, another important challenge is the need to compensate for the differential optical path differences across a telescope array - differences nearly as large as the telescope separations. For very long baselines, the classical method of free-space propagation through optical trombones would require mechanisms to optical tolerance on a topographically challenging scale – and in vacuum.
 
Fortunately, in recent years, developments in quantum information have led to proposals for alternate technologies which may support extremely long telescope baselines, and hence ultra-high angular resolution.

\section{Astronomy's Quest for Optimized Imaging}

Astronomers have always sought better image quality from their observatories, whether with improved optics, superior observatory locations, or with technology (automated guiding, adaptive optics). Beyond improvements in traditional stigmatic imaging, there have also been initiatives directed toward enhanced extraction of image content by other means, such as shift-and-add, speckle imaging, bispectrum analysis, and related post-processing methods.

Astronomers have in some cases utilized macroscopic optics in an attempt to optimize telescopes and arrays for special purpose measurements, usually on a somewhat ad hoc basis.  Early applications of interferometry for measurement of double stars and stellar diameters is reviewed by McAlister \cite{McAlister2020}. Conventional optical interferometry with multiptelescope arrays proved to be a powerful general purpose tool for astronomy, though with available angular resolution only slightly advanced over the last two decades. For high contrast imaging with a single aperture, analysis of the apodization of a telescope pupil has been used to approach the theoretical limit for a broad class of single aperture coronagraphs\cite{Guyon2006}. For nulling of a bright star and detection of nearby exoplanets, telescope arrays with optimized (de)phasing has been extensively explored analytically (e.g.~\citenum{Bracewell1978}), with some implementations. Aperture masking in various forms (e.g.~\citenum{Huby2012}) have brought further options for pupil manipulation and extraction of image parameters. 

With the powerful tools of quantum optics and quantum information theory, telescopic imaging can potentially be systematically optimized, for both single apertures and for multi-telescope arrays~\cite{Tsang2016, Lee2022, Grace2022, Cosmo2020, Sajjad2024, Padilla2024}. Historically, optical array layouts have been based on intuition, tradition and limited modeling. Insights and analysis from quantum information theory may support optimization of the geometry of an optical array as well as the choice of beam-combining and measurement strategies, both for general purpose imaging and in the quest for more specifically defined image information, including super-resolution. An accessible, fully self-contained introduction to some of these quantum information theory methods for readers in the astronomy community has been provided in~\cite{Sajjad2024} with a detailed treatment of several types of multiple-telescope systems.

\section{Bridging the classical to the quantum world in astronomical interferometry}

\begin{figure}[t]
    \includegraphics[width=0.99\linewidth]{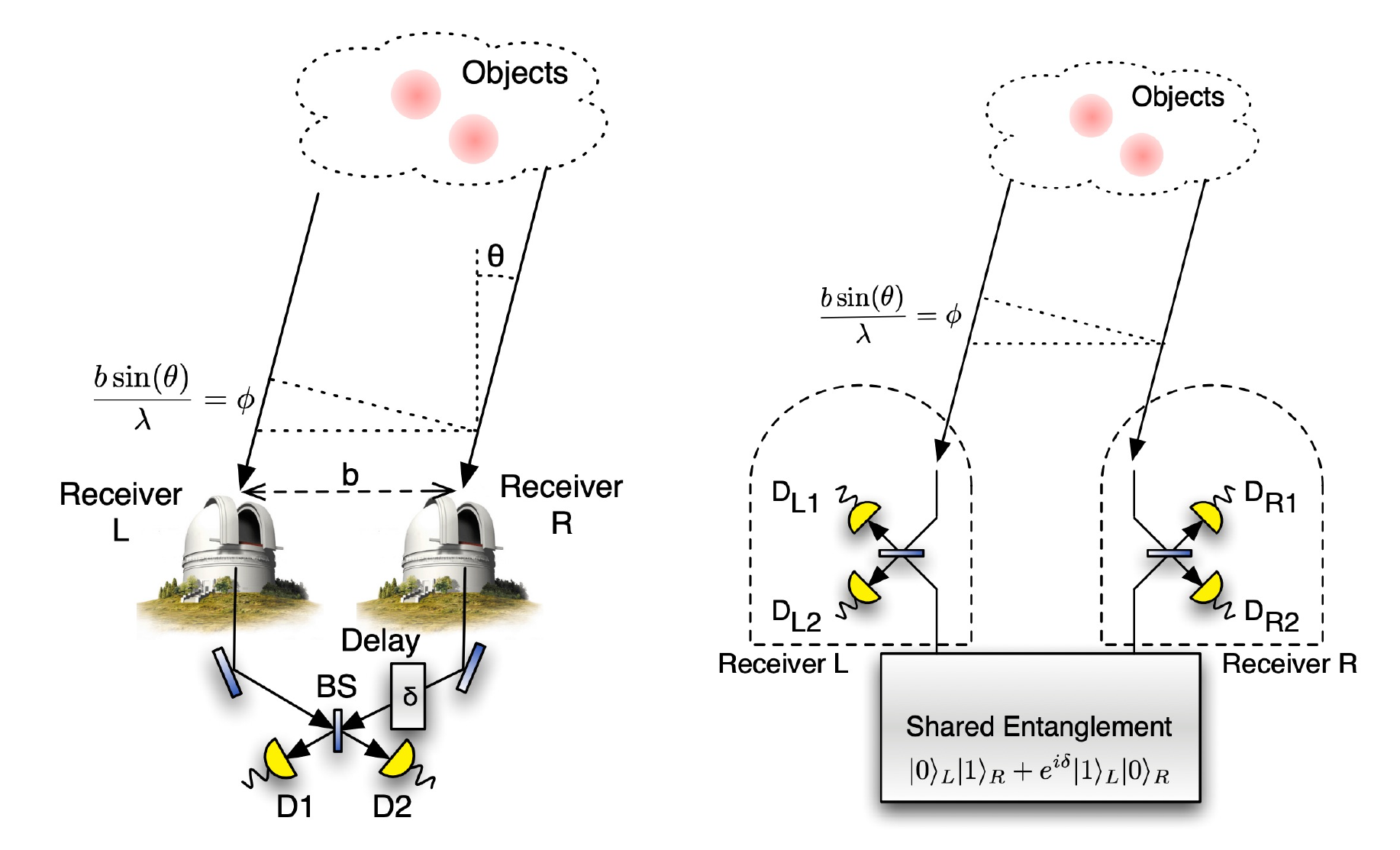}
    \caption{(Left) Classical ``direct detection'' interferometry schematic revisited by \citenum{Gottesman2012}. (Right) Schematic of the Gottesman protocol\cite{Gottesman2012} where an interference measurement of an astronomical photon is made by creating entangled states at both telescope sites. } 
    \label{fig:Fig1}
\end{figure}

\subsection{Overcoming the transmission loss over large optical baselines}

Photon occupation numbers for thermal emission in the optical band from sources such as stars are very small.  
As a result, the rate of incoming photons captured by even a large telescope aperture is not high enough to survive significant transmission by current state-of-the-art optical fiber to distances much greater than a few km. 
Direct beam combination via single mode (SM) optical fibers may be possible, in a limited spectral range, for up to several 10’s of kilometers. Near 1550 nm, the most efficient SM fibers have losses as low as 0.22 dB/km (FS.com Inc). This corresponds to throughput of $8\%$ over 50 km, which may be plausible, and less than $1\%$ over 100 km, which we believe is insufficient for facility-scale operation. Note that this idealized throughput does not take into account various intrinsic and extrinsic losses such as injection and coupler efficiency and those due to mechanical stress and fiber fabrication tolerances. Other wavelengths have significantly poorer performance.

In order to mitigate the transmission loss inherent in the conventional “direct detection” interferometry, Gottesman et  al.\ (2012)\cite{Gottesman2012} proposed a novel method utilizing shared quantum entanglement. The ``Gottesman protocol'' is best introduced by first recasting the incoming astronomical light in a classical ``direct detection" interferometry scenario from a quantum-mechanics perspective. 
Fig.~\ref{fig:Fig1} (Left) represents direct detection from \citenum{Gottesman2012}, where light from two separated apertures is brought together through free space or optical fibers to a beam-combiner. 
Because of the sparse photon number, we can assume that each spatio-temporal mode has 1 photon at most. The superposition of the astronomical photon over the two apertures can then be described by the state

\begin{equation}
|0\rangle_L|1\rangle_R + e^{i\phi}|1\rangle_L|0\rangle_R,
\label{eq:1}
\end{equation}

\noindent
where $|0\rangle$ and $|1\rangle$ represent 0 and 1-photon states, \(L\) and \(R\) are the two apertures, and \(\phi\) is the relative phase shift. The beams are combined in a beam splitter (BS) after imposing a delay ($\delta$) in one arm introduced by a variable delay line. The complex visibility is then measured from the two BS outputs, where the photon detection probabilities at detectors D1 and D2 (Fig.~\ref{fig:Fig1}) are:

\begin{equation}
 \frac{1 + \Re (V e^{-i\delta})}{2}\,\,\text{and}\,\, \frac{1 - \Re (V e^{-i\delta})}{2},
\label{eq:2}
\end{equation}

\noindent
respectively. By sweeping through $\delta$, both the amplitude and phase of the visibility ($V$) can be recovered.

\subsection{Gottesman Protocol}

In the Gottesman protocol\cite{Gottesman2012} (Fig.~\ref{fig:Fig1}, Right), an entangled state is distributed to the two apertures from a central entangled photon source with a variable delay introduced on one arm. Two interference measurements are then conducted, one at each site. The resulting photon ‘clicks’ detected are then post-selected for correlated and anti-correlated BS outputs of one photon arriving at each site, with  probabilities as in Eqn.~\ref{eq:2}, to once again recover the visibility. In effect, this is equivalent to teleportation of the astronomical photon state between the apertures using the lab-prepared entangled state as the channel and circumvents the physical transmission of the signal from the astronomical photon. Instead, entangled states are sent to the two apertures. Moreover the delay compensation can be inserted in the path of the entangled pair instead of the astronomical photon.
However, in the Gottesman protocol, $50\%$ of the signal is lost when the “lab” photon and the astronomical photon arrive at the same site. 

A major technological barrier to realizing this two-photon interference protocol is the fact that we need to maintain a steady and reliable supply of continuously refreshed entangled states to perform correlation measurements between the two sites to sample every mode (including the vacuum states) to catch the rare occupied mode. The entangled states must be well-matched in all modes (arrival time, frequency, bandwidth,  and polarization) to completely interfere with the incoming astronomical photon and give a good visibility measurement with strong correlation and anti-correlation statistics. Moreover, entangled photon generation at such a rate (i.e.~$>10$ GHz\cite{Gottesman2012,Khab2019prl}) is extremely daunting given that current single-photon sources that can generate entangled states have achieved to-date a maximal rate of the order of 100 kHz (e.g., \citenum{Bouillard2019}).

\begin{figure}[t]
    \centering
    \includegraphics[width=0.8\linewidth]{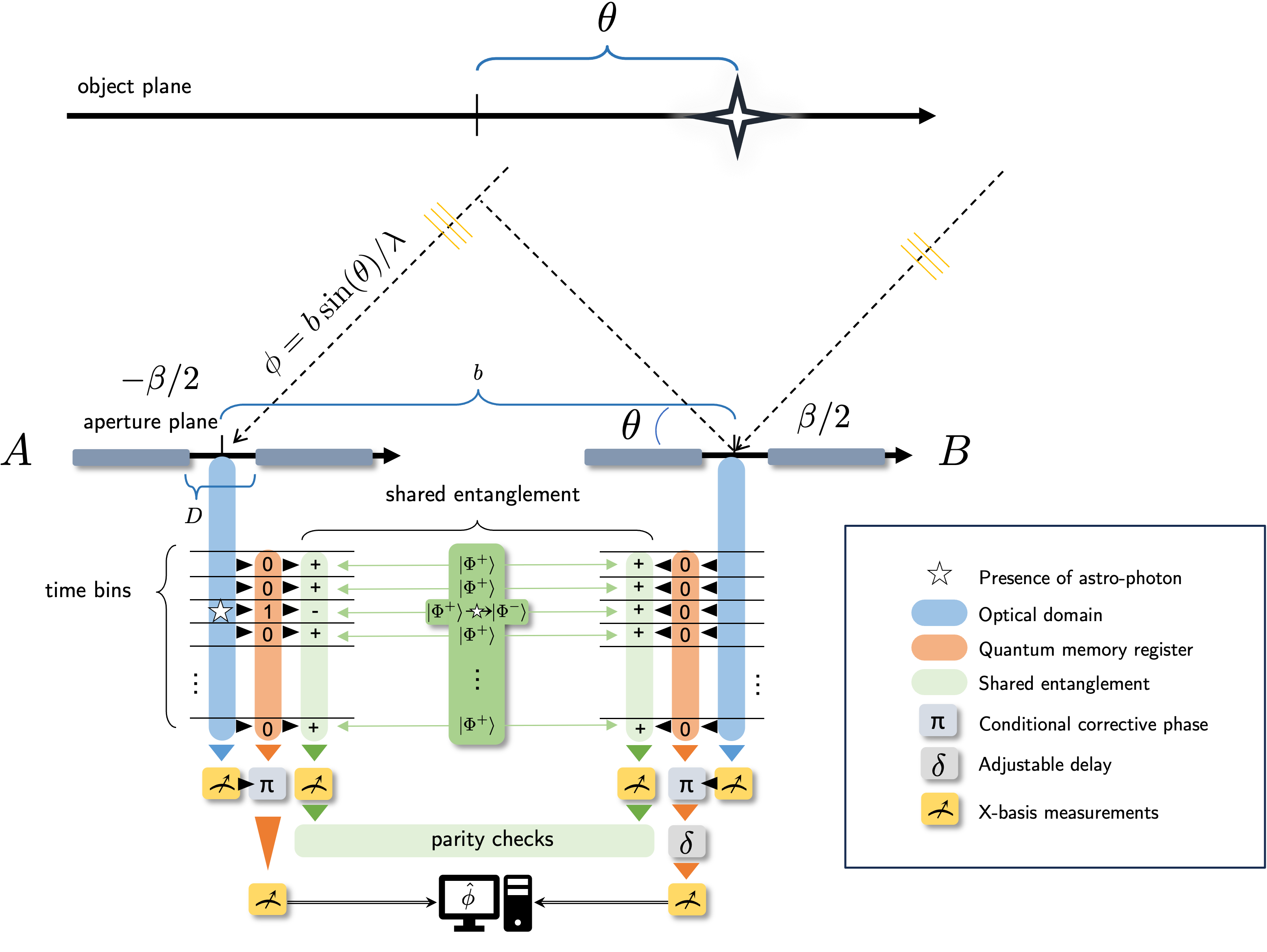}
    \caption{Schematic of the Khabiboulline protocol\cite{Khab2019prl} (with no binary encoding).  The optical channel is represented in blue, the memory qubits in orange and the entangled states in green. The photon is denoted by a  star (in the third bin). The “meter” symbols represent measurements. The parity check reveals an odd parity for the third memory pair, identifying the filled photon temporal mode. Local measurements on the astronomical photons (blue channel) are relayed classically and the relative phase, $\phi$, is derived from measurements on the memory qubits, effectively “teleporting” the astronomical photon state between the sites using the entanglement.} 
    \label{fig:Fig2}
\end{figure}

\subsection{Khabiboulline Protocol}

A significant step towards alleviating the Gottesman protocol's extremely large required rate of entangled photon pair generation was proposed by Khabiboulline et  al.\ (2019)\cite{Khab2019prl}. 
In this protocol (Fig.~\ref{fig:Fig2}), quantum memories are utilized to mitigate the requirement on continually refreshed entanglement distribution. In quantum memories, each memory “qubit” is a superposition of two states (e.g.~ 1, 0 or + ,  -.) and could be implemented as electron spin states of an atom, photons in optical cavities,  SiV defects in diamond, etc. The interaction with the incoming photon both imprints the bin number (arrival time) in binary code and the photon state on the memory qubits at each site.  A  ‘parity check’ on the memory qubits, using entangled states distributed to the two sites,  identifies the time-bin that has interacted with the incoming photon — without revealing which aperture or site is involved. If $M$ memory qubits are required to sample one photon on average, this binary encoding reduces the number of memory qubits and entangled states needed to $\sim$ log$_2(M+1)$, thus overcoming the serious problem of the extremely large entanglement rates in the Gottesman protocol\cite{Gottesman2012}.  In the final step, the states of the astronomical photons between baselines can be  “teleported” via the entanglement channel and the relative phase determined,  equivalent to one direct interference measurement to calculate the visibility. The Khabiboulline protocol exploits the low photon number to limit the rate of entangled states to the photon arrival rate rather than the optical bandwidth used.  Unlike the Gottesman protocol, this protocol does not discard half the astronomical photons received by the telescope array. 
 
The Gottesman and Khabiboulline protocols are the two main quantum protocols that have been presented to-date to overcome the distance-limitation of classical interferometry via distributed quantum entanglement. 
In a follow-up paper\cite{Khab2019pra}, Khabiboulline et al.\ notably extend their protocol and describe a conceptual path to utilize multiple apertures and frequency channels with both pair-wise and multi-aperture interferometry.
Another recent work~\cite{Padilla2024} presents a generalized theoretical framework that builds upon these quantum protocols and incorporates techniques such as spatial mode sorting in an interferometric telescope array to achieve the angular resolution limit defined by quantum information theory\cite{Tsang2016,Sajjad2024,Padilla2024}.

%Another recent work~\cite{Padilla2024} extends the Khabiboulline et al. protocol by describing the entanglement-based version of a general receiver in which we have a mode sorter at each telescope, and each spatial mode is then transported  through a single mode fiber to a central location where all the signals are combined in an arbitrary linear interferometer with any given unitary transformation. Using analytical tools from quantum information theory ~\cite{Sajjad2024}, it is possible to come up with the optimal unitary transformation for mixing the various spatial modes from the different locations for estimating a general parameter associated with the scene.

%Another recent work~\cite{Padilla2024} presents a generalized theoretical framework that builds upon these quantum protocols and incorporates techniques such as spatial mode sorting in an interferometric telescope array to achieve the angular resolution limit defined by quantum information theory \cite{Tsang2016,Sajjad2024,padilla2024}.

% The overall idea is that the paper describes a way to obtain the entanglement-based equivalent of a system that transports the light collected at the different telescopes to a central location and mixes it in an arbitrary form (i.e. with an arbitrary unitary transformation) before measuring it. And then we can leverage the tools from quantum information theory to come up with the optimal way to mix the various signals for estimating the parameter of interest in the scene.

\subsection{Addressing Fringe Tracking and OPD Compensation at Very Long Baselines}

Matching the path delay between the two arms (to within the coherence length) is a major challenge for conventional interferometry. Indeed, using delay lines of the order of tens of kilometers verges on the impossible given mechanical and terrain constraints.
Although this has not been explicitly addressed by the Khabiboulline protocol\cite{Khab2019prl}, quantum entanglement offers a way forward. Since the bulk of the OPD is deterministic, it should be possible to compensate for the OPD by timing offsets. For example, if we assume a 1 microsecond delay between photon arrival at the two sites, the parity checking could be done between the appropriate time-bins (memory qubits) after offsetting the timing of the parity check between sites by 1 microsecond. We are exploring whether this is feasible and practical.
We note that the Gottesmann protocol explicitly puts the delay in one arm of the ``lab'' entangled pair to compensate for the large geometric OPD.

Fringe tracking and correcting for fast, small phase errors remains an issue. As presented by \citenum{Khab2019prl}, fringe tracking will remain a classically-employed method, along with adaptive optics. However, enough photon signal must arrive over atmospheric timescales in order for these essential techniques to remain possible. 
We note here that well-known techniques like baseline bootstrapping and closure phases are equally applicable with entanglement-based methods and can be used for fringe tracking and phase recovery.

Using timing delays for rapid, small phase measurements for fringe tracking and visibility measurements will need to be of the order of 100s of ps timing precision and accuracy, and places extreme timing requirements that will have to be supplied by next-generation atomic clock distributed networks.

Timing delays also require the ability for the memories to hold the information for up to ms intervals. 
Today's quantum memory cells have lifetimes (decoherence time) on the order of microseconds, but in some cases up to 2 ms for photonic qubits. This is a further area that needs to be improved as this lifetime must encompass not only the capturing of wavefronts, but also all the quantum computing operational steps that lead up to the final interference measurement.

Without multiple wavelength channels, the local qubit memory cell requirements in the Khabiboulline protocol are of the order of \(\sim 20-30\) qubits, and an entanglement supply rate of \(\sim 200 \, \text{kHz}\) for the photon arrival rate for a 10th magnitude star, collecting area of 10 square meters and detector bandwidth of 10 GHz. Both of these technical requirements seem feasible to achieve in the near-future given today's capabilities.

We acknowledge exploration of timing requirements is the tip of the iceberg in defining the full set of technical requirements to perform astronomical observations with quantum-enhanced techniques. In a path towards an on-sky demonstration (Sec.~\ref{sec:onsky}), we plan a further investigation of requirements and feasibility including a signal-to-noise-ratio performance estimate. 

\section{Laboratory Demonstrations of Entanglement and Quantum Memory}

With current quantum technology, the Gottesman protocol\cite{Gottesman2012} has a higher readiness level compared to the Khabiboulline protocol\cite{Khab2019prl} since it does not require high-fidelity quantum memories. Brown et al.\ (2023, and these proceedings)\cite{Brown2023} have demonstrated a table-top proof-of-principle experiment for the Gottesman protocol by implementing an entangled photon source working as the shared entanglement resource and an illuminated double slit as the target. In their experiment, the entangled photon pair at a central wavelength of 830 nm with a 10 nm bandwidth is generated by pumping the type-II spontaneous parametric down-conversion (SPDC) process of a potassium dihydrogen phosphate (KDP) crystal. After passing through individually tunable optical path delays, two generated entangled photons are separated, and then sent through fibers. One of these serves as the heralding signal and the other interferes with the photon collected by the two fiber collimators (the apertures). The optical path delays are set to compensate for the relative delay in photon arrival time at the two apertures. Their experimental results show that the Gottesman protocol can reconstruct the wavefront phase based on the interference fringes, collected at two remote sites, by consuming the entangled photon pairs. 

However, the optical fibers still incur a significant loss and add noise to entangled photons when a longer storage time is needed in practice, which diminishes the rate and fidelity of sharing entanglement. To circumvent this issue, researchers have considered storing entanglement in quantum memories with higher fidelity and lifetime than photons transmitted in fibers. With plenty of quantum memory nodes at each telescope site, the entanglement resources can be distributed continuously and retrieved immediately when needed. During the past decade, distributing bipartite quantum entanglement through optical channels between remote quantum memory nodes has been demonstrated among multiple platforms, including trapped ions, neutral atoms, quantum dots, and color centers in diamonds. Indeed, there are still many engineering challenges awaiting. For example, these quantum memories usually operate at different frequencies than the source photons, either laboratory or from astronomical objects. Since the interference must occur between photons at the same frequency, frequency conversion is necessary. 

With the emerging practical solutions to these challenges, exciting and practical demonstrations have been exhibited recently. In one of the most recent field tests, Knaut et  al.\ (2024)\cite{Knaut2024} demonstrated the entanglement generation between two $\mathrm{Si}^{29}$-vacancy-in-diamond quantum memories apart by 40 km low-loss fiber through the Boston area. The Bell entangled state they distributed among two remote nuclear spins has an initial fidelity of 0.77 and can be preserved above 0.5 for 500 ms. The generation rate is less than 1 pair per second. Although the rate and fidelity are lower than needed, this experiment is still a significant step toward implementing the Khabiboulline protocol for quantum-enhanced long-baseline interferometry. 
%\Com{You may want to pull some figures from their papers, but I'm not very sure about the policy of doing that in a conference paper.} 

\section{Concepts for First On-sky Demonstration of Indirect Beam Combination} 
\label{sec:onsky}

Certainly the quantum technologies needed for astronomy will be first fully developed in the laboratory with simulated sources. Nevertheless, we regard an end-to-end demonstration with stellar photons to be an essential step in the path to a possible major development initiative of the future.
 
The laboratory infrastructure required for quantum communication experiments strongly encourages us to investigate ways to bring the sky to the laboratory. In fact, many modern astronomy instruments exploit such a configuration, often utilizing optical fibers to convey carefully ``tailored'' starlight to a stable and controlled environment.  Adaptive optics ensures a high-quality point spread function. Optical fibers, which in the most demanding applications are single mode, carry the light from the relatively hostile environment of the rotating telescope in the open air to a more friendly and stable location where large and sensitive optical systems can be installed and maintained. This is the approach that we envision for on-sky demonstrations and operation. An observatory ``system'' will deliver spatially coherent stellar photons of well defined properties (wavelength, polarization, spatial mode), with near-zero and adjustable optical path difference between the channels.
 
We distinguish here two stages of demonstration:

\begin{itemize}
  \item \textbf{Stage 1: Implementing the Gottesman Protocol\cite{Gottesman2012}} The first stage is demonstration of interferometry with entanglement-supported beam combination, where quantum communication substitutes for direct combination of the stellar light. Optical path differences, however, are controlled by existing conventional methods.
  \item \textbf{Stage 2: Implementing the Khabiboulline Protocol\cite{Khab2019prl}} The second stage demonstration, utilizing quantum memory, would add the capability of compensating for optical path differences by controlled time delay in the quantum measurement.
\end{itemize}
 
Further, each stage of demonstration is likely to involve two possible phases.

\begin{itemize}
\item \textit{Phase A.} The first and lowest overhead phase would be a relatively simple demonstration with a single aperture telescope, sub-divided to supply two separate spatial samples of the stellar wavefront. This program would possibly be supported in part by NOIRLab and possibly at a national facility. At Kitt Peak in Arizona, we have identified two NSF-owned telescopes, currently unutilized, each of which appears to offer  potential venues for the first phase experiments. These telescopes could potentially supply spatial wavefront samples with apertures up to 40 cm in diameter and with separations up to 80 – 130 cm.

\item \textit{Phase B.} The second phase might be carried out at an existing astronomical optical array, possibly at the CHARA Array with support of the CHARA/GSU team.  CHARA can provide stellar light from 2 or more adaptive optics corrected 1.0 m telescopes, with separations in the range 30-330 meters. The existing instrument complement can provide stellar photons to a laboratory environment via single mode optical fibers with optical path difference fully compensated and controlled. A demonstration of quantum beam combination in this venue offers the opportunity to explore a broad range of quantum physics innovations in beam combination and detection, in a realistic environment, with direct comparison to classical methods, and the potential to carry out novel astronomical measurements.
\end{itemize}

\section{Bridging the Astronomy and Quantum Communities}

In pursuit of exploring the applications of quantum-enhanced methods for interferometry, new collaborations and initiatives are being seeded to bridge the astronomy and quantum information science communities. A direct example of these joint community efforts was the 1-day workshop ’Charting Quantum Horizons: Establishing a Roadmap for Microarcsecond Astronomy,’ which was organised by NSF NOIRLab in collaboration with the Arizona Quantum Initiative, the Center for Quantum Networks (CQN), and CHARA (See \citenum{Lau2024} for a meeting report).  The workshop attracted over 80 in-person participants, two-thirds of which had an astronomy background and the remaining third had a background in quantum information science. Notably, graduate students accounted for over a fifth of the registered participants, which underscored the interest and future prospects for implementing quantum-enhanced techniques to astronomical interferometry. One of the major takeaways from the workshop was a consensus between the  astronomy and quantum communities that a near-term ($\sim$5 year), on-sky demonstration of quantum-enhanced interferometry is an important next step. Given the enthusiasm from both astronomers and quantum information scientists, NSF NOIRLab is launching a new ``Quantum Telescope Initiative'' to facilitate the development of a quantum-enhanced telescope.

\section{Summary of Advantages and Challenges for Quantum-Enhanced Interferometry}

We conclude this work with a summary of the advantages and the challenges for applying quantum-enhanced techniques for astronomical interferometry.

\noindent
\textbf{Advantages of Quantum Interferometry}

\begin{itemize}
    \item Transmission losses are significantly reduced or even avoided.
    \item Large optical path difference can be compensated using time delays or offsets.
    \item It may be possible to filter out incoherent photons (noise) taking advantage of the narrow bandwidths and synchronization of the memory qubits.
\end{itemize}

\noindent
\textbf{Important Unknowns and Limitations}

\begin{itemize}
    \item Generation at high rates of high-fidelity entangled states.
    \item Narrow effective bandwidths and repeat rates.
    \item Timing precision and accuracy of loading the incoming astronomical photon onto the quantum memory (must be significantly less than phase measurement requirements).
    \item Synchronization.
    \item Electrical/photonic control of raster scanning incoming astronomical photon to quantum shift register.
    \item Many of the protocols, e.g., X-basis measurements on a single rail (1 for photon, 0 for vacuum) state photon qubit, are not fully explored for physical implementation.
    \item Errors in the quantum gates and measurements.
    \item Small photon numbers and susceptibility to noise.
\end{itemize}

From a broader perspective, we note that quantum networks and computing are active areas of research to realize mature quantum networks. Significant resources are being invested on topics such as reliable quantum repeaters, heralded photons, entanglement distillation and quantum error correction, and quantum memory.
The field of astronomy therefore has much to gain by engaging with the field of quantum information science and exploring the utility of quantum techniques to achieve breakthrough science.

\acknowledgments
We thank Aziza Suleymanzade, Brian Smith, Pieter-Jan Stas, John Monnier, and Matthew Brown for the valuable interactions and feedback on this work.
The work by JKR, RML, STR, BM, SC, MR, and FR is supported by NOIRLab, which is managed by the Association of Universities for Research in Astronomy (AURA) under a cooperative agreement with the U.S. National Science Foundation.
Supported by the international Gemini Observatory, a program of NSF NOIRLab, which is managed by the Association of Universities for Research in Astronomy (AURA) under a cooperative agreement with the U.S. National Science Foundation, on behalf of the Gemini partnership of Argentina, Brazil, Canada, Chile, the Republic of Korea, and the United States of America.
JGR is supported by the NASA Space Technology Graduate Research Opportunity under Grant No. 80NSSC23K1211.
Additionally, we acknowledge contributions of authors affiliated with the Engineering Research Center for Quantum Networks (CQN), awarded by the NSF and DoE under cooperative agreement number 1941583, for synergistic research support.

% References
\bibliography{report} % bibliography data in report.bib
\bibliographystyle{spiebib} % makes bibtex use spiebib.bst

\end{document}